# Minimax Linear Estimation at a Boundary Point

Wayne Yuan Gao*

June 15, 2017

**Abstract**

This paper characterizes the minimax linear estimator of the value of an unknown function at a boundary point of its domain in a Gaussian white noise model under the restriction that the first-order derivative of the unknown function is Lipschitz continuous (the second-order Hölder class). The result is then applied to construct the minimax optimal estimator for the regression discontinuity design model, where the parameter of interest involves function values at boundary points.

*Gao: Department of Economics, Yale University, 28 Hillhouse Ave., New Haven, CT 06511, USA. Email: wayne.gao@yale.edu. The author would like to thank Tim Armstrong for guidance and comments.

# 1 Introduction

This paper studies the problem of finding the minimax linear estimator of the value of an unknown function $f$ at a boundary point, where the function $f$ is known to lie in some smoothness class of functions $\mathscr{F}$. This paper focuses on the Hölder class of the second order, i.e., $\mathscr{F}$ is taken to be the space of continuously differentiable function whose first-order derivative is Lipschitz continuous with Lipschitz constant $C$. A Gaussian (or white-noise) regression model is considered, where a stochastic process $\{Y(t) : t \in \mathbb{R}_+ \equiv [0, \infty)\}$ is observed and assumed to be generated according to

$$dY(t) = f(t) \cdot dt + \sigma \cdot dW(t),$$

with $W$ being a standard Brownian motion (or Weiner) process. The parameter of interest is taken to be $Lf := f(0)$, the value of $f$ at the boundary point of its domain, and we seek to obtain a linear estimator $\hat{L}$ for $f(0)$ that minimizes the worst-case mean squared error, i.e., $\hat{L}$ solves $\inf_{\hat{L} \in \mathscr{L}} \sup_{f \in \mathscr{F}} \mathbb{E}_f \left[ \left( \hat{L} - Lf \right)^2 \right]$. As shown in Donoho [1994], the problem of linear minimax estimation is essentially equivalent to solving the modulus problem, or to find the least-favorable function $f^*$. Following Zhao [1997], it can be shown that this is again equivalent to minimizing $\|f\|^2$ subject to $f(0) = b$ in $\mathscr{F}$.

Zhao [1997] considers a similar problem but takes the domain of the unknown function $f$ to be $\mathbb{R}$, so that the parameter of interest $f(0)$ is the value of $f$ at an interior point of $\mathbb{R}$. As a result, by the strict convexity of $\|\cdot\|^2$, the solution $f^*$ must be symmetric about 0, which implies that $f^{*'}(0) = 0$, imposing an initial condition on $f^{*'}$ in addition to $f^*(0) = b$. Zhao [1997] then provides a characterization of the solution $f^*$.

The key difference of this paper with Zhao [1997] lies in that in this paper $f(0)$ is the value of $f$ at the boundary point of the domain of $f$. As a result, the convexity of $\|\cdot\|^2$ no longer imposes any condition on the first-order derivative of $f$, and the least favorable function $f^*$ requires the optimal specification of $f^{*'}(0)$. This paper provides a complete characterization of the solution $f^*$, which has two main features. First, the Lipschitz continuity constraint on $f^{*'}$ is binding on $\mathbb{R}_+$: $\left|f^{*''}\right| = C$ on $\mathbb{R}_+$ (almost everywhere). Second, without loss of generality assuming $f^*(0) > 0$, the optimal initial condition on the first-order derivative satisfies $f^{*'}(0) < 0$.

This paper complements Zhao [1997] and a line of other researches on the solutions to the minimax estimation problems under various settings[1], such as Taikov [1968], Legostaeva and Shiryaev [1971], Sacks and Ylvisaker [1978], Sacks and Ylvisaker [1981], Li [1982], Silverman [1984], Wahba [1990], Fan [1993], Korostelev [1994], Cheng, Fan, and Marron [1997], Fan, Gasser, Gijbels, Brockmann, and Engel [1997],

---

[1]The various settings studied by the cited works differ on which smoothness class (Hölder class, Taylor class, Sobolev class, etc.) the unkonwn function is restricted to, whether the regression model is continuous or discrete, and whether the point of interest lies in the interior or at the boundary.

Leonov [1999] and Cai and Low [2003] . See Armstrong and Kolesár [2016a,b] for a review of these results along with the construction of honest confidence intervals for nonparametric regression models based on minimax optimal kernels.

A particular application of the result is to estimate the regression-discontinuity parameter in the regression-discontinuity (RD) design model, which, introduced first by Thistlethwaite and Campbell [1960], arises frequently in many economic scenarios. Imbens and Lemieux [2008] provide a review of various settings of the RD design model and a practical guide for estimation based on the work of Cheng, Fan, and Marron [1997], who show the minimax optimality of their method in the Taylor class. As noted in Sacks and Ylvisaker [1981] and Zhao [1997], often times the Hölder class is a more natural restriction to make than the Taylor class[2]. Consider the following Gaussian nonparametric regression model:

$$\dot{Y}_t = f(t) + \sigma \dot{W}_t, \quad W_t \sim BM(1)$$

where $f$ is a function defined on $\mathbb{R}$. Suppose that there is a regression discontinuity at 0, i.e.,

$$Lf := \lim_{x \searrow 0} f(x) - \lim_{x \nearrow 0} f(x) \neq 0.$$

We might be interested in estimating the RD parameter $Lf$: if a "treatment" is assigned whenever $x \geq 0$, then the parameter of interest $Lf$ captures the "treatment effect" at $x = 0$. Define $f_+, f_- : \mathbb{R}_+ \to \mathbb{R}$ by

$$f_+(x) := \begin{cases} f(x), & x > 0 \\ \lim_{y \searrow 0} f(y), & x = 0 \end{cases}, \quad f_-(x) := \begin{cases} f(-x), & x > 0 \\ \lim_{y \searrow 0} f(-y), & x = 0 \end{cases}$$

The RD parameter $Lf = f_+(0) - f_-(0)$ is then the difference between $f_+$ and $f_-$ at the boundary point of their domain $\mathbb{R}_+$. In section 3, we demonstrate how the main result of this paper can be applied to construct the minimax linear estimator for the RD parameter.

The paper is organized as follows. Section 2 presents the model and the main result. Section 3 considers in more details the application to the RD design model. Section 4 concludes the paper. The Appendix contains all proofs.

## 2   Model and Main Result

Following Donoho [1994], consider the general model

$$y = Kf + \epsilon \tag{1}$$

---

[2]Note that the Hölder class imposes a uniform bound on a derivative of a certain order globally, while the Taylor class only bounds the derivative locally at a point. The global restriction allowed by the Hölder class, when empirically reasonable, helps improve the minimax optimality of estimation.

where $y, \epsilon \in \mathscr{Y}$, a Hilbert space with inner product $\langle \cdot, \cdot \rangle$ and norm $\|\cdot\|$, and $\epsilon$ is standard Gaussian with respect to $\langle \cdot, \cdot \rangle$, i.e., $\forall g \in \mathscr{Y}$, $\langle g, \epsilon \rangle \sim N\left(0, \|g\|^2 \sigma^2\right)$ with $\sigma^2 > 0$ known. $f \in \mathscr{F}$ is a unknown real function, where $\mathscr{F}$ is the set of admissible functions to which we restrict our attention, and $K : \mathscr{F} \to \mathscr{Y}$ is a known linear operator on $\mathscr{F}$. Let $L : \mathscr{F} \to \mathbb{R}$ be a given functional, and we are interested in estimating the parameter of interest $Lf$.

Following Zhao [1997], this paper considers the special case of Gaussian (or white noise) regression model, in which we observe $\{Y(t) : t \in [0, \infty)\}$ with

$$dY(t) = f(t) \cdot dt + \sigma \cdot dW(t), \tag{2}$$

where $W(t)$ is a standard Brownian motion (or Wiener) process, $\sigma$ is a known parameter and $f$ is an unknown function in $\mathscr{F}$. Equivalently, we specialize the general model (1) by taking

$$y = \dot{Y}, \quad Kf = f, \quad \epsilon = \sigma \dot{W}$$

and $\mathscr{Y} = L_2(\mathbb{R}_+)$ with the standard inner product

$$\langle g, h \rangle := \int_0^\infty g(t) h(t) dt \quad \forall g, h \in L_2(\mathbb{R}_+).$$

By the standard Gaussianity of $\epsilon$, $\forall g \in \mathscr{Y}$,

$$\langle g, \epsilon \rangle = \sigma \int_0^\infty g(t) dW(t) \sim N\left(0, \sigma^2 \|g\|^2\right).$$

Moreover, we restrict $f$ to be continuously differentiable and $f'$ to be Lipschitz continuous with parameter $C$, i.e,

$$\mathscr{F} = \mathscr{F}_H(2, C) = \left\{ g \in L_2(\mathbb{R}_+) \cap C^1(\mathbb{R}_+) : \left|g'(t) - g'(s)\right| \leq C |t - s| \ \forall t, s \in \mathbb{R}_+ \right\}.$$

where $\mathscr{F}_H(\gamma, C)$, as defined in Lepski and Tsybakov [2000], denotes the Hölder class with order $\gamma > 0$ and constant $C > 0$:

$$\mathscr{F}_H(\gamma, C) := \left\{ g \in L_2(\mathbb{R}_+) \cap C^{\lfloor \gamma \rfloor}(\mathbb{R}_+) : \left|g^{(\lfloor \gamma \rfloor)}(t) - g^{(\lfloor \gamma \rfloor)}(s)\right| \leq C |t - s|^{\gamma - \lfloor \gamma \rfloor} \ \forall t, s \in \mathbb{R}_+ \right\}$$

where $\lfloor \gamma \rfloor := \max \{k \in \mathbb{N} : k < \gamma\}$.

We consider the parameter of interest

$$Lf = f(0),$$

i.e., the value of $f$ at the boundary point of its domain $[0, \infty)$. The distinguishing feature of this paper relative to Zhao [1997] lies in that in this paper the parameter of interest is the value of $f$ at a boundary point of its domain, while Zhao [1997] considers the value of $f$ at an interior point of its domain. The fact that 0 is a boundary point complicates the analysis relative to the interior case, as the constraint imposed by Lipschitz continuity becomes slack at the boundary point, leaving arbitrariness in the "initial condition" on $f'$, which will be clarified in the subsequent analysis.

Let $\mathscr{L}$ denote the space of linear estimators of $f(0)$:

$$\mathscr{L} := \left\{ \hat{L} = \langle \psi, y \rangle \equiv \int_0^\infty \psi(t)\, dY(t) \quad \text{for some function } \psi \in L_2(\mathbb{R}_+) \right\}.$$

Our goal is then to find a linear minimax estimator $\hat{L} \in \mathscr{L}$ of $f(0)$, or equivalently to solve

$$R^*_{\mathscr{L}}(\mathscr{F}) = \inf_{\hat{L} \in \mathscr{L}} \sup_{f \in \mathscr{F}} \mathbb{E}_f \left[ \left( \hat{L} - Lf \right)^2 \right]. \tag{3}$$

By Ibragimov and Khas'minskii [1985], as $\mathscr{F}$ is convex, closed and centrosymmetric,

$$R^*_{\mathscr{L}}(\mathscr{F}) = \sup_{f \in \mathscr{F}} \frac{\sigma^2 (Lf)^2}{\sigma^2 + \|f\|^2} \tag{4}$$

and, given a solution $f^* \in \mathscr{F}$ to (4), the linear minimax estimator $\hat{L}^*$ can then be obtained as

$$\hat{L}^* = \int_0^\infty \psi^*(t)\, dY(t) \tag{5}$$

$$\psi^*(t) = \frac{f^*(0)}{\sigma^2 + \|f^*\|^2} \cdot f^*(t) \tag{6}$$

This transforms the optimization problem (3) over $\hat{L} \in \mathscr{L}$ to an equivalent problem (4) over $\mathscr{F}$ to search for the least favorable function $f^*$.

To solve (4), we first observe that

$$\sup_{f \in \mathscr{F}} \frac{\sigma^2 (Lf)^2}{\sigma^2 + \|f\|^2} = \sup_{\delta > 0} \frac{\sigma^2 \cdot b^2(\delta)}{\sigma^2 + \delta^2} \tag{7}$$

where

$$b(\delta) := \sup_{f \in \mathscr{F},\ \|f\| = \delta} |f(0)|. \tag{8}$$

If, for each $\delta$, we can solve (8) and obtain the value $b(\delta)$ as well as the optimizer $f^*_\delta$, then we can solve (7) for the optimal $\delta^*$ and set $f^* = f^*_{\delta^*}$, with which we can obtain $\hat{L}^*$ by (5) and (6).

Hence, the key problem is to solve (8), and we proceed by solving its dual problem:

$$\inf_{f \in \mathscr{F},\ |f(0)| = b} \|f\|^2 = \int_0^\infty f^2(t)\, dt \tag{9}$$

i.e., the least favorable $f_b$ is the one with the minimal norm given the "initial condition" $f(0) = \pm b$ and the smoothness constraint imposed by the Lipschitz-continuity condition on $f'$. Without loss of generality, let $f(0) = b$. We use convex optimization theory in Hilbert space to establish existence and uniqueness. The proof constructs a closed and bounded set, which, being weakly compact, suffices for the attainment of the minimum of a lower semi-continuous convex function.

**Theorem 1.** *There exists a unique (up to the indistinguishability under $\|\cdot\|$) function $f_b^*$ that solves $\inf_{f \in \mathscr{F},\ f(0)=b} \|f\|^2$.*

With the existence and uniqueness of the solution to problem (9) ensured by Theorem 1, we proceed to provide an explicit characterization of the solution. To simplify the problem, notice that Lemma 1 of Zhao [1997] still applies.

**Lemma 1.** *Let $f_{b,C}^*$ denote the unique solution to problem (9), and $I_{b,C}^* := \left\|f_{b,C}^*\right\|^2$. Then:*

$$I_{b,C} = \frac{b^{5/2}}{\sqrt{C}} \cdot I_{1,1}$$

$$f_{b,C}(t) = b \cdot f_{1,1}\left(\sqrt{\frac{C}{b}} x\right)$$

Notice that $f \in \mathscr{F} = \mathscr{F}_H(2, C)$ implies that $|f''(t)| \leq C$ for any $t \in \mathbb{R}_+$ at which $f''$ is well-defined. By the density of $C^2(\mathbb{R}_+)$ in $C^1(\mathbb{R}_+)$, as well Lemma 1, it then suffices to solve the following problem:

$$\inf_{\substack{f(0)=1 \\ |f''| \leq 1}} \int_0^\infty f^2(t)\,dt \tag{10}$$

**Lemma 2** (Theorem 3 of Zhao (1997)). *Let $f_0^*$ denote the solution to problem (10) subject to the constraint that $f'(0) = 0$. Then, $\forall t \in \mathbb{R}_+$,*

$$f_0^*(t) = 1 + \int_0^t \int_0^\tau \left(\sum_{k=1}^\infty (-1)^k \mathbf{1}\{s \in [a_{k-1}, a_k)\}\right) ds\,d\tau \tag{11}$$

*with $a_0 = 0$, $a_1 = k_0^*$, $a_k = a_{k-1} + k_0^*\left(1 + \sqrt{k_0^{*2} - 1}\right)(k_0^{*2} - 1)^{\frac{k-2}{2}}$ for all $k \geq 2$, where $k_0^* \approx 1.02889$ is the solution to*

$$\min_{k_0 \geq 0} \frac{\frac{23}{30}k_0^5 - 2k_0^3 + 2k_0}{1 - \sqrt{(k_0^2 - 1)^5}}.$$

*Moreover, $I_0^* := \|f_0^*\|^2 \approx 0.76402$.*

In this paper, we are interested in characterizing $f^*$, the solution to problem (10) without the initial constraint on $f'(0)$ in Zhao [1997]. By Lemma 1 and Lemma 2, if $f^{*'}(t_0) = 0$ for some $t_0 \in \mathbb{R}_+$, then $f^*|_{[t_0,\infty)}$ can be easily obtained by a shifting and rescaling of $f_0^*$.

Specifically, $\forall y \in (-\infty, 1)$, consider

$$g_y^*(t) := \begin{cases} y + \frac{1}{2}\left(t - \sqrt{2(1-y)}\right)^2, & t \in \left[0, \sqrt{2(1-y)}\right) \\ y \cdot f_0^*\left(\frac{t - \sqrt{2(1-y)}}{\sqrt{|y|}}\right), & t \in \left[\sqrt{2(1-y)}, \infty\right) \end{cases} \quad (12)$$

where $f_0^*$ is defined in (11). Notice that $g_y^*(0) = 1$ and $\left|g_y^{*''}\right| \leq 1 \ \forall y \in (-\infty, 1)$, so $g_y^* \in \mathscr{F}_1$. In particular, note that

$$g_0^*(t) = \frac{1}{2}\left(t - \sqrt{2}\right)^2 \cdot \mathbf{1}\left\{t \in \left[0, \sqrt{2}\right)\right\} \geq 0$$

and $\|g_0^*\|^2 = \frac{\sqrt{2}}{5} \approx 0.28284 < I_0^*$.

**Lemma 3.** $\forall y \in (-\infty, 1)$, if $f^*(t) \geq y$ for all $t \in \left[0, \sqrt{2(1-y)}\right]$, then $f^*(t) \geq g_y^*(t)$ for all $t \in \left[0, \sqrt{2(1-y)}\right]$.

**Theorem 2.** $f^* = g_y^*$ for some $y \in (-\infty, 1]$.

Given Theorem 2, problem (10) is then equivalent to solve for the optimal $y \in (-\infty, 1)$:

$$\min_{y \in (-\infty, 1)} \|g_y^*\|^2 = \int_0^{\sqrt{2(1-y)}} \left[y + \frac{1}{2}\left(t - \sqrt{2(1-y)}\right)^2\right]^2 dt + |y|^{\frac{5}{2}} \cdot I_0^*$$

$$= \frac{1}{15}\sqrt{2(1-y)}\left(3 + 4y + 8y^2\right) + I_0^* \cdot |y|^{\frac{5}{2}} \quad (13)$$

**Theorem 3.** There exists a unique solution $y^* < 0$ to problem (13) with

$$y^* \approx -0.12455$$
$$I^* = \|g_{y^*}^*\|^2 \approx 0.26672.$$

The unique solution $f^*$ to problem (10) is given by $f^* = g_{y^*}^*$. Moreover:

- The corresponding initial condition on $f^{*'}$ is given by
$$f^{*'}(0) = -\sqrt{2(1-y^*)} \approx -1.4997.$$

- $f^*$ has bounded support $[0, \bar{t}]$ with
$$\bar{t} := \sqrt{2(1-y^*)} + \sqrt{-y^*} \cdot \left(k_0^* + \frac{1 + \sqrt{k_0^{*2} - 1}}{1 - \sqrt{k_0^{*2} - 1}}\right) \approx 2.44121.$$

Optimization problem (13) is solved numerically using Mathematica. $\bar{t}$ is obtained by Lemma 1 and the Corollary in Zhao [1997].

With Theorem 3, we may now recover the solution to problem (3) according to the following steps.

**Corollary 1.** *The unique solution $f^*_{b,C}$ to and the value $I^*_{b,C}$ of problem (9) are given by*

$$f^*_{b,C}(t) = b \cdot f^*\left(\sqrt{\frac{C}{b}} t\right) \quad \forall t \in \mathbb{R}_+, \quad I^*_{b,C} = \frac{b^{5/2}}{\sqrt{C}} \cdot I^*.$$

Let $\delta^2 = I^*_{b,C} = \frac{b^{5/2}}{\sqrt{C}} \cdot I^*$, we obtain

$$b(\delta) = C^{\frac{1}{5}} I^{*-\frac{2}{5}} \cdot \delta^{\frac{4}{5}}. \tag{14}$$

Plugging (14) into (7), we solve

$$\sup_{\delta > 0} \frac{\sigma^2 \cdot C^{\frac{2}{5}} I^{*-\frac{4}{5}} \cdot \delta^{\frac{8}{5}}}{\sigma^2 + \delta^2} \quad \Rightarrow \quad \delta^* = 2\sigma. \tag{15}$$

Now we obtain the linear minimax estimator $\hat{L}^*$ as

$$\psi^*(t) = \frac{b(\delta^*)}{\sigma^2 + \delta^{*2}} \cdot f^*_{b(\delta^*),C}(t)$$

$$= \frac{2^{\frac{8}{5}} \cdot I^{*-\frac{4}{5}}}{5} C^{\frac{2}{5}} \sigma^{-\frac{2}{5}} \cdot f^*\left(\left(\frac{I^* C^2}{4\sigma^2}\right)^{\frac{1}{5}} \cdot t\right)$$

$$\hat{L}^* = \int_0^\infty \psi^*(t) \, dY(t)$$

and the linear minimax risk is given by

$$R^*_{\mathscr{L}}(\mathscr{F}) = \frac{2^{\frac{8}{5}} \cdot I^{*-\frac{4}{5}}}{5} C^{\frac{2}{5}} \sigma^{\frac{8}{5}} \approx 8.72575 \times C^{\frac{2}{5}} \sigma^{\frac{8}{5}}.$$

# 3 Application to the Regression Discontinuity Design Model

We now apply the main result to the sharp regression discontinuity (RD) design model.

Consider the following Gaussian nonparametric regression model:

$$\dot{Y}_t = f(t) + \sigma \dot{W}_t, \quad W_t \sim BM(1) \tag{16}$$

where $f$ is a function defined on $\mathbb{R}$. Given $f$, define $f_+, f_- : \mathbb{R}_+ \to \mathbb{R}$ by

$$f_+(x) := \begin{cases} f(x), & x > 0 \\ \lim_{y \downarrow 0} f(y), & x = 0 \end{cases}, \quad f_-(x) := \begin{cases} f(-x), & x > 0 \\ \lim_{y \uparrow 0} f(y), & x = 0 \end{cases}$$

We restrict $f_+, f_-$ to lie in the smooth class $\mathscr{F}_H(2,C)$. Equivalently, we assume that

$$f \in \mathscr{F} = \{g : \mathbb{R} \to \mathbb{R} \text{ s.t. } g_+, g_- \in \mathscr{F}_H(2,C)\}.$$

We are interested in estimating the regression-discontinuity parameter

$$L_{RD} f := f_+(0) - f_-(0).$$

By (3) and (4), the minimax risk is given by

$$I^*_{RD,b,C} = \sup_{f \in \mathscr{F}} \frac{\sigma^2 (L_{RD} f)^2}{\sigma^2 + \|f\|^2} = \sup_{\delta > 0} \frac{\sigma^2 \cdot b_{RD}^2(\delta)}{\sigma^2 + \delta^2}$$

where

$$b_{RD}(\delta) := \sup_{f \in \mathscr{F},\ \|f\| = \delta} |f_+(0) - f_-(0)|.$$

Again we proceed by solving the dual problem

$$\inf_{f \in \mathscr{F},\ |f_+(0) - f_-(0)| = b} \|f\|^2 = \|f_+\|^2 + \|f_-\|^2, \tag{17}$$

and its solution is characterized by the following:

**Theorem 4.** *The solution $f^{RD}$ to (17) is an odd function given by*

$$f^{RD}_+ = -f^{RD}_- = f^*_{b/2,C},$$

*where $f^*_{b/2,C}$ is defined in Corollary 1.*

*The value of problem (17) is given by*

$$I^*_{RD,b,C} = 2 I^*_{b/2,C} = \frac{b^{5/2}}{2^{3/2} \sqrt{C}} \cdot I^*.$$

Solving $\delta_{RD}^2 = I^*_{RD,b,C}$, we obtain $b_{RD}(\delta) = 2^{\frac{3}{5}} C^{\frac{1}{5}} I^{*-\frac{2}{5}} \delta^{\frac{4}{5}}$. Clearly $\delta^*_{RD} = 2\sigma$ as before. Therefore, the optimal kernel is given by

$$\psi^*_{RD}(t) = -\psi^*_{RD}(-t) = \frac{b(\delta^*)}{\sigma^2 + \delta^{*2}} \cdot f^*_{b(\delta^*)/2,C}(t)$$

$$= \frac{2^{\frac{9}{5}} I^{*-\frac{4}{5}}}{5} C^{\frac{2}{5}} \sigma^{-\frac{2}{5}} \cdot f^* \left( \left( \frac{I^* C^2}{2\sigma^2} \right)^{\frac{1}{5}} \cdot t \right), \quad \forall t \in \mathbb{R}_+$$

and the minimax linear estimator for the RD parameter is given by

$$\hat{L}_{RD}^* = \int_{-\infty}^{0} \left(-\psi_{RD}^*\left(-t\right)\right) dY\left(t\right) + \int_{0}^{\infty} \psi_{RD}^*\left(t\right) dY\left(t\right)$$

$$= \int_{0}^{\infty} \psi_{RD}^*\left(t\right) dY\left(t\right) - \int_{0}^{\infty} \psi_{RD}^*\left(t\right) dY\left(-t\right).$$

This suggests that we may separately estimate $f_+(0)$ and $f_-(0)$ using the kernel $\psi_{RD}^*$, and then calculate the difference to obtain the minimax optimal estimator $f_+(0) - f_-(0)$.

## 4 Conclusion

Building on Zhao [1997], this paper characterizes the minimax linear estimator of the value of an unknown function at a boundary point of its domain in a Gaussian white noise model under the restriction that the first-order derivative of the unknown function is Lipschitz continuous. We show that the main result of this paper can be applied to the regression-discontinuity design model, a popular model in which the parameter of interest involves function values at boundary points.

## Appendix

### Proof of Theorem 1

*Proof.* Clearly $\|\cdot\|^2 : \mathscr{F} \to \mathbb{R}$ is continuous (and thus lower semi-continuous), and it is strongly convex: $\forall f, g \in \mathscr{F}$ s.t. $\|f - g\| \neq 0$ and $\alpha \in (0, 1)$,

$$\|\alpha f + (1-\alpha) g\|^2 = \int \left(\alpha f(t) + (1-\alpha) g(t)\right)^2 dt$$

$$< \int \left(\alpha f^2(t) + (1-\alpha) g^2(t)\right) dt$$

$$= \alpha \|f\|^2 + (1-\alpha) \|g\|^2.$$

The set $\mathscr{F}_b = \{f \in \mathscr{F} : f(0) = b\}$ is clearly closed in $\mathscr{F}$. Fix any $g \in \mathscr{F}_b$, let

$$\mathscr{F}_{b,g} := \{f \in \mathscr{F}_b : \|f\| \leq \|g\|\}.$$

$\mathscr{F}_{b,g}$ is clearly closed, bounded and convex subset of $\mathscr{F}$. Note also that $\mathscr{F} = \mathscr{F}_H(2, C)$ is a closed and convex subspace of a Hilbert space, and it is thus reflexive. Then, by a mathematical theorem on convex optimization in reflexive Banach space[3], a lower semi-continuous convex function on a reflexive Banach space attains its minimum value on every bounded, closed and convex subset of the Banach space. Hence, there exists a $f_b^* \in \mathscr{F}_{b,g}$ that $\|\cdot\|^2$ attains its minimum on $\mathscr{F}_{b,g}$ at $f_b^*$. By the definition of $\mathscr{F}_{b,g}$, $f_b^*$ is also a minimizer for $\|\cdot\|^2$ on $\mathscr{F}_b$. By the strict convexity of $\|\cdot\|^2$, $f_b^*$ must be unique. $\square$

---
[3]Theorem 2.11 in Barbu and Precupanu [2012], p. 72.

## Proof of Lemma 3

*Proof.* Suppose that $f^*(t_0) < g_y^*(t_0)$ for some $t_0 \in \left(0, \sqrt{2(1-y)}\right)$. Consider the set

$$T_1 := \{t \in [0, t_0] : f^*(t) = g_0^*(t)\}.$$

Clearly, $T_1$ is nonempty (as $0 \in T_1$) and closed by the continuity of $f^*$ and $g_y^*$. Hence, $t_1 := \max T_1$ is well-defined, and $f^*(t) < g_y^*(t)$ for $t \in (t_1, t_0]$. Then it must be that $f^{*'}(t_1) \leq g_y^{*'}(t_1) = \frac{1}{2}\left(t_1 - \sqrt{2(1-y)}\right)$. Otherwise, $f^{*'}(t_1) > g_y^{*'}(t_1)$ and $f^*(t_1) = g_y^*(t_1)$ together imply that $f^*(t) > g_y^*(t)$ for some $t \in (t_1, t_0]$. Moreover, we must have $f^{*'}(t) < g_y^{*'}(t)$ on $(t_1, t_1']$ for some $t_1' \in (t_1, t_0)$; otherwise it is not possible to have $f^*(t) < g_y^*(t)$ for $t \in (t_1, t_0]$. As a result,

$$f^*\left(\sqrt{2(1-y)}\right) = f^*(t_1) + \int_{t_1}^{t_1'} f^{*'}(t)\,dt + \int_{t_1'}^{\sqrt{2}} \left[f^{*'}\left(t_1'\right) + \int_{t_1'}^{\tau} f^{*''}(\tau)\,d\tau\right] dt$$

$$< g_y^*(t_1) + \int_{t_1}^{t_1'} g_0^{*'}(t)\,dt + \int_{t_1'}^{\sqrt{2}} \left[g_0^{*'}\left(t_1'\right) + \int_{t_1'}^{\tau} g^{*''}(\tau)\,d\tau\right] dt$$

$$= g_y^*\left(\sqrt{2(1-y)}\right) = y,$$

contradicting the supposition that $f^* \geq y$ on $\left[0, \sqrt{2(1-y)}\right]$. $\square$

## Proof of Theorem 2

*Proof.* We consider two possible cases separately.

*Case 1:* $f^* \geq 0$ *on* $\mathbb{R}_+$.
Let $y = 0$ and consider $g_0^*$. By Lemma 3, $f^* \geq g_0^*$ on $\left(0, \sqrt{2}\right)$. As $g_0^*(t) = 0$ on $\left[\sqrt{2}, \infty\right)$, we have $f^* \geq g_0^* \geq 0$ on $\mathbb{R}_+$.
Now, suppose that $f^*(t_0) > g(t_0)$ for some $t_0 \in \left(0, \sqrt{2}\right)$. Then $f^*(t) > g(t) > 0$ on $(t_0 - \epsilon, t_0 + \epsilon) \subseteq \left(0, \sqrt{2}\right)$ for some $\epsilon > 0$. Then,

$$\|f^*\|^2 > \|g_0^*\|^2$$

contradicting the supposition that $f^*$ is a minimizer of $\|\cdot\|^2$ on $\mathscr{F}_1$.

Hence, it must be that $f^* = g_0^*$ and thus $\|f^*\|^2 = \|g_0^*\|^2 = \frac{\sqrt{2}}{5} \approx 0.28284$.

*Case 2:* $f^*(t_0) < 0$ *for some* $t_0 \in (0, \infty)$.
Let $T_0$ denote the set of critical points of $f^*$ on $\mathbb{R}_+$:

$$T_0 := \left\{ t \in \mathbb{R}_+ : f^{*'}(t) = 0 \right\}$$

Notice that $T_0$ is nonempty. Suppose to the contrary that $T_0 = \emptyset$. Then there cannot be any local minimum. As a result, $t_0 = \arg\min_{t \in [0, t_0]} f^*(t)$ and $t = \arg\min_{\tau \in [t_0, t]} f^*(\tau)$ $\forall t \in (t_0, \infty)$; otherwise there will be a local minimum. Then $f^*(s) \leq f^*(t_0) < 0$, and thus $\|f^*\| = \infty$, contradicting the supposition that $f^*$ is a minimizer of $\|\cdot\|^2$ on $\mathscr{F}_1$.

Moreover, as $f^{*'}$ is continuous, $T_0$ is closed. Hence, $t_1 := \min T_0$ is well-defined. Let

$$y := f^*(t_1).$$

(i) Suppose $y = 0$. Then it must be that $f^*(t) > 0$ on $(0, t_1)$ and, by the optimality of $f^*$, $f^*(t) = 0$ for all $t \geq t_1$. But this implies that $f^* \geq 0$ on $\mathbb{R}_+$, contradicting the supposition that $f^*(t_0) < 0$ for some $t_0 \in (0, \infty)$.

(ii) Suppose $y \geq 1$. Then $f^{*'} > 0$ and $f^* \geq f^*(0) = 1$ on $[0, t_1]$. Hence,

$$\|f^*\|^2 \geq t_1^2 + y^{\frac{5}{2}} \cdot I_0^* \geq I_0^* > \|g_0^*\|^2$$

contradicting the optimality of $f^*$ in $\mathscr{F}_1$.

(iii) Suppose $y \in (0, 1)$. Then $f^{*'} < 0$ and $f^*(t) \geq f^*(t_1) = y$ on $[0, t_1]$. We then show that $t_1 = \sqrt{2(1-y)}$. If $t_1 < \sqrt{2(1-y)}$,

$$f^*(0) = f^*(t_1) - \int_0^{t_1} \left[ 0 - \int_\tau^{t_1} f^{*''}(s) \, ds \right] d\tau$$
$$\leq y + \frac{1}{2} t_1^2 < y + \frac{1}{2} \cdot 2(1-y)$$
$$= 1$$

contradicting the supposition that $f^* \in \mathscr{F}_1$. If $t_1 > \sqrt{2(1-y)}$, we have $f^* \geq y$ on $[0, t_1] \supseteq \left[0, \sqrt{2(1-y)}\right]$. By Lemma 3, we then have $f^*(t) \geq g_y^*(t) > 0$ for all $t \in \left[0, \sqrt{2(1-y)}\right]$. Therefore,

$$\|f^*\|^2 = \left( \int_0^{\sqrt{2(1-y)}} + \int_{\sqrt{2(1-y)}}^{t_1} + \int_0^{\sqrt{2(1-y)}} \right) f^{*2}(t) \, dt$$
$$\geq \int_0^{\sqrt{2(1-y)}} g_y^{*2}(t) \, dt + \left( t_1 - \sqrt{2(1-y)} \right) \cdot y + y^{\frac{5}{2}} I_0^*$$
$$> \|g_y^*\|^2$$

contradicting the optimality of $f^*$ again. Hence, we must have $t_1 = \sqrt{2(1-y)}$. As $f^*(0) = 1$ is possible only if $f^* = g_y^*$ on $[0, t_1]$, we conclude that $f^* = g_y^*$ on $\mathbb{R}_+$.

(iv) Suppose $y < 0$. By exactly the same argument as in (iii), we conclude that $t_1 \geq \sqrt{2(1-y)}$. Now suppose that $t_1 > \sqrt{2(1-y)}$. By Lemma 3, we again have $f^* \geq g_y^*$ on $\left[0, \sqrt{2(1-y)}\right]$. As $f^{*\prime} < 0$ on $[0, t_1]$, there exists a unique $z \in (0, t_1)$ such that $f^*(z) = 0$. Note that $g_y^*\left(\sqrt{2(1-y)} - \sqrt{-2y}\right) = 0$ at $\sqrt{2(1-y)} - \sqrt{-2y} \in \left(0, \sqrt{2(1-y)}\right)$, so $f^*\left(\sqrt{2(1-y)} - \sqrt{-2y}\right) \geq 0$ and thus

$$z \geq \sqrt{2(1-y)} - \sqrt{-2y}, \tag{18}$$

Hence, as $f^* \geq g_y^* \geq 0$ on $\left[0, \sqrt{2(1-y)} - \sqrt{-2y}\right]$, we deduce that

$$\int_0^{\sqrt{2(1-y)} - \sqrt{-2y}} f^{*2}(t)\, dt \geq \int_0^{\sqrt{2(1-y)} - \sqrt{-2y}} g_y^{*2}(t)\, dt \tag{19}$$

Let $h(t) := y + \frac{1}{2}(t - t_1)^2$. Notice that, $\forall t \in [0, t_1]$,

$$f^*(t) = f^*(t_1) - \int_0^{t_1} \left[0 - \int_\tau^{t_1} f^{*\prime\prime}(s)\, ds\right] d\tau$$

$$\leq y + \frac{1}{2}(t - t_1)^2 = h(t),$$

so $f^*\left(t_1 - \sqrt{-2y}\right) \leq h\left(t_1 - \sqrt{-2y}\right) = 0$, and thus

$$z \leq t_1 - \sqrt{-2y}. \tag{20}$$

Hence, as $f^* \leq h_y^* \leq 0$ on $\left[t_1 - \sqrt{-2y}, t_1\right]$, we deduce that

$$\int_{t_1 - \sqrt{-2y}}^{t_1} f^{*2}(t)\, dt \geq \int_{t_1 - \sqrt{-2y}}^{t_1} h_y^{*2}(t)\, dt$$

$$= \int_{-\sqrt{-2y}}^0 \left(y + \frac{1}{2}t^2\right)^2 dt$$

$$= \int_{\sqrt{2(1-y)} - \sqrt{-2y}}^{\sqrt{2(1-y)}} g_y^{*2}(t)\, dt \tag{21}$$

By (18) - (21), when $t_1 > \sqrt{2(1-y)}$, either (18) or (21) holds with strict inequality, so have

$$\|f^*\|^2 = \left(\int_0^z + \int_z^{t_1} + \int_{t_1}^\infty\right) f^{*2}(t)\, dt$$

$$> \left( \int_0^{\sqrt{2(1-y)}-\sqrt{-2y}} + \int_{t_1-\sqrt{-2y}}^{t_1} \right) f^{*2}(t)\,dt + |y|^{\frac{5}{2}} \cdot I_0^*$$

$$\geq \left( \int_0^{\sqrt{2(1-y)}-\sqrt{-2y}} + \int_{\sqrt{2(1-y)}-\sqrt{-2y}}^{\sqrt{2(1-y)}} \right) g_y^{*2}(t)\,dt + |y|^{\frac{5}{2}} \cdot I_0^*$$

$$= \|g_y^*\|^2,$$

contradicting the optimality of $f^*$ in $\mathscr{F}_1$. Hence, we again deduce that $t_1 = \sqrt{2(1-y)}$ and $f^* = g_y^*$ on $\mathbb{R}_+$. $\qquad\square$

## Proof of Theorem 4

*Proof.* Without loss of generality, assume that $f_+(0) = a$ and $f_-(0) = a - b$. Given $f_+(0), f_-(0)$, we may directly apply Corollary 1 and conclude that

$$\inf_{f_+ \in \mathscr{F}_H(2,C), f_+(0)=a} \|f_+\|^2 = I^*_{|a|,C} = \frac{|a|^{5/2}}{\sqrt{C}} \cdot I^*,$$

with the minimum attained by $f_+ = f^*_{|a|,C}$. Similarly,

$$\inf_{f_- \in \mathscr{F}_H(2,C), f_-(0)=a-b} \|f_+\|^2 = I^*_{|a-b|,C} = \frac{|a-b|^{5/2}}{\sqrt{C}} \cdot I^*,$$

with the minimum attained by $f_- = f^*_{|a-b|,C}$.

$$\inf_{f \in \mathscr{F}, f_+(0)=a, f_-(0)=a-b} \|f\|^2 = \left( |a|^{5/2} + |a-b|^{5/2} \right) \cdot \frac{I^*}{\sqrt{C}}.$$

Then problem (17) may be transformed to problem (22):

$$\inf_{f \in \mathscr{F}, |f_+(0)-f_-(0)|=b} \|f\|^2 = \inf_{a \in \mathbb{R}} \inf_{f \in \mathscr{F}, f_+(0)=a, f_-(0)=a-b} \|f\|^2$$

$$= \frac{I^*}{\sqrt{C}} \cdot \inf_{a \in \mathbb{R}} \left( |a|^{5/2} + |a-b|^{5/2} \right). \tag{22}$$

As $h(x) := x^{5/2}$ is a convex function on $\mathbb{R}_+$, we have

$$\frac{1}{2}\left( |a|^{5/2} + |a-b|^{5/2} \right) \geq \left( \frac{1}{2}|a| + \frac{1}{2}|a-b| \right)^{5/2}$$

where the equality holds if and only if $|a| = |a-b|$, i.e. $a = \frac{b}{2}$. Hence,

$$\inf_{f \in \mathscr{F}, |f_+(0)-f_-(0)|=b} \|f\|^2 = \frac{I^*}{2^{3/2}\sqrt{C}} \cdot |b|^{5/2}$$

with the minimum attained by $f^{RD}$ such that

$$f_+^{RD} = -f_-^{RD} = f^*_{b/2,C}.$$

$\qquad\square$